\documentclass[journal]{IEEEtran}

\ifCLASSINFOpdf

\else

\fi
\usepackage{graphicx}
\usepackage{amsmath,amssymb,amsfonts}
\usepackage{multirow}
\usepackage{hyperref}
\usepackage{epstopdf}
\usepackage{booktabs}
\usepackage{algorithm}
\usepackage{algpseudocode}
\usepackage{amsmath}

\usepackage{etoolbox}
\makeatletter
\patchcmd{\@makecaption}
  {\scshape}
  {}
  {}
  {}
\makeatletter
\patchcmd{\@makecaption}
  {\\}
  {:\ }
  {}
  {}
\makeatother
\def\tablename{Table}

\begin{document}

\title{\huge Brain Structure-Function Fusing Representation Learning using Adversarial Decomposed-VAE for Analyzing MCI}


\author{Qiankun Zuo, Baiying Lei,  Ning Zhong, Yi Pan, Shuqiang Wang

\thanks{Qiankun Zuo, Yi Pan and Shuqiang Wang are with the Shenzhen Institutes of Advanced Technology, Chinese Academy of Sciences, Shenzhen, 518055, China.}
\thanks{Baiying Lei is with the School of Biomedical Engineering, Shenzhen University, Shenzhen, 518060, China}
\thanks{Ning Zhong is with Department of Systems Life Engineering, Maebashi Institute of Technology, Maebashi 371-0816, Japan}
}

\markboth{}%
{Shell \MakeLowercase{\textit{et al.}}: Bare Demo of IEEEtran.cls for IEEE Journals}
%

\maketitle

\begin{abstract}
Integrating the brain structural and functional connectivity features is of great significance in both exploring brain science and analyzing cognitive impairment clinically. However, it remains a challenge to effectively fuse structural and functional features in exploring the brain network. In this paper, a novel brain structure-function fusing-representation learning (BSFL) model is proposed to effectively learn fused representation from diffusion tensor imaging (DTI) and resting-state functional magnetic resonance imaging (fMRI) for mild cognitive impairment (MCI) analysis. Specifically, the decomposition-fusion framework is developed to first decompose the feature space into the union of the uniform and the unique spaces for each modality, and then adaptively fuse the decomposed features to learn MCI-related representation. Moreover, a knowledge-aware transformer module is designed to automatically capture local and global connectivity features throughout the brain. Also, a uniform-unique contrastive loss is further devised to make the decomposition more effective and enhance the complementarity of structural and functional features. The extensive experiments demonstrate that the proposed model achieves better performance than other competitive methods in predicting and analyzing MCI. More importantly, the proposed model could be a potential tool for reconstructing unified brain networks and predicting abnormal connections during the degenerative processes in MCI.
\end{abstract}


%
\IEEEpeerreviewmaketitle

\section{Introduction}
%
%
%
%
\IEEEPARstart{M}{ild cognitive impairment} (MCI) is considered an early stage of Alzheimer's Disease (AD) among older people~\cite{ref_01}. It is characterized by memory loss, aphasia, and other brain function decline~\cite{ref_02}. Although not all older adults with MCI will develop AD, the annual conversion rate is 10\%-15\%. As the stage of AD is irreversible and incurable, early cognitive training and rehabilitation treatment are the keys to delaying or preventing the onset of dementia. Therefore, it is essential to develop effective methods for the diagnosis of MCI~\cite{ref_03, ref_04, ref_05}.

The brain network is suitable for characterizing the structural or functional relationships between brain regions by diffusion tensor imaging (DTI) or resting-state functional magnetic resonance imaging (fMRI). As parts of the brain's structural or functional connections may alter in people with MCI~\cite{ref_06, ref_07}, it is common to extract connectivity-based features for early cognitive disease detection, which captures more effective topological information beyond Euclidean space~\cite{hu2021point,lu2016,you2022brain,hu2022srt}. The general way of describing these features is to first split the whole brain into several spatially distributed regions of interest (ROIs) and then compute the connection strength between each other from imaging data~\cite{wang2021image,jing2022ta}. Previous studies~\cite{ref_08,ref_09} extracted the connectivity-based features from unimodal data and then built a classifier for cognitive disease detection. Since neuroimages from different modalities carry complementary information, current works~\cite{ref_10,ref_11} mainly focus on multimodal fusion by graph convolutional network (GCN) and have achieved superior performance in disease diagnosis. However, these works heavily depend on the structural connectivity by empirical methods, which may lead to a large error in connection strength calculation because of the manually different parameter settings in certain software toolboxes. It may lose much valuable information for disease prediction. Besides, the high noise and changing connectives derived from fMRI make it difficult to fuse with DTI, which cannot fully capture the complex brain network features in cognitive disease analysis.

As the transformer network~\cite{ref_11.5} has an excellent ability in capturing global information and model longer-distance dependencies for image recognition~\cite{ref_12,ref_13}, it is more suitable to automatically extract structural features of the spatially distributed brain ROIs and determine the connection strength among them. Since the common and complementary information from unimodal data is often mixed together, the multimodal fusion effect has been dramatically improved by decomposition-fusion representation learning via variational autoencoder (VAE) in disease analysis~\cite{ref_14,ref_15}. However, these works focus on image feature extraction in Euclidean space and ignore the representation learning in topological space. It cannot analyze the connectivity features among spatial distributed ROIs in brain disease prediction. Moreover, the decomposed representations need to be adaptively integrated to learn effective connectivity features for disease analysis.

Inspired by the above observations, in this paper, a novel model termed brain structure-function fusing-representation learning (BSFL) is proposed to generate unified brain networks for predicting abnormal brain connections based on fMRI and DTI. Specifically, the knowledge-aware transformer network is designed to extract structural features for each ROI from DTI. Then the structural features and the functional features extracted from fMRI are sent to the decomposed variational graph autoencoders to decompose the feature space into uniform and unique spaces representing the common and complementary information for each modality. After that, the decomposed representations are utilized to reconstruct the input features to retain the unimodal information. Meanwhile, the representation-fusing generator combines these representations and generates unified brain networks, which are sent to the dual discriminator to make them class-discriminative and distribution consistent. To ensure the effectiveness of decomposition, a uniform-unique contrastive loss function is utilized to constrain the distance in the decomposed representations within each modality and between modalities. As a result, the unified connectivity-based features are obtained to fully capture MCI-related information and provide reliable analysis of brain network abnormalities. The main contributions of this framework are as follows:
\begin{itemize}
	\item The novel BSFL model is proposed to first learn the uniform and unique representations of each modality in topological spaces, and then adaptively integrate them to generate unified brain networks. It can greatly enhance the structural-functional feature fusion and effectively recognize the connectivity features that are highly related to MCI.
	\item The uniform-unique contrastive loss is devised to maximize the distance of the uniform and unique representations within each modality and minimize the distance of uniform representations between modalities, which makes the decomposition more effective and enhances the complementarity of structural and functional features.
	\item The knowledge-aware transformer (KAT) is designed to extract brain region features from DTI by introducing knowledge of the brain parcellation atlas. The proposed KAT can automatically learn the local and global connectivity features and capture the MCI-related structural information.
\end{itemize}

The rest of this paper is organized as follows. The related works are briefly described in Section II. The details of the proposed model are presented in Section III. In Section IV, the proposed BSFL and other competing methods are compared, and experimental results are presented on the public database. The reliability of the experimental results and the limitations of the proposed model are discussed in Section V. Finally, Section VI concludes the remarks of this study.

\section{Related Work}
The current brain network analysis methods for cognitive disease can be summarized in three categories: structural connectivity-based, functional connectivity-based, and multimodal connectivity-based approaches. The first approach focuses on morphology or water diffusion information to extract interrelated features of predefined ROIs for AD analysis. He et al.~\cite{ref_17} utilized cortical thickness information from T1-MRI to construct brain networks to analyze the abnormal topology property between patient groups and healthy controls. Similarly, Wang et al.~\cite{ref_18} constructed structural brain networks from DTI data to evaluate graph topological coefficients and demonstrated that the AD group had decreased global efficiency and local efficiency compared with normal controls. The brain network can also be characterized by the neural activity measured from each brain region. The second approach constructed functional connectivity from fMRI or Electroencephalogram (EEG) and built classifiers to diagnose early AD. The work in ~\cite{ref_19} investigated subgroups of functional connectivities using EEG data and found abnormal changes in hub regions in AD patients. By defining spatial distributed ROIs, Jie et al.~\cite{ref_20} utilized fMRI to extract connectivity-based features with multiple ROIs, which improved the MCI diagnosis accuracy and provided valuable biomarkers for treatment. Considering multimodal data provides complementary information, the third approach jointly used structural and functional connectivity to construct unified brain networks for AD diagnosis and treatment. Song et al.~\cite{ref_11} fused functional and structural connectivity-based features by applying adaptive calibrated GCN to boost AD prediction accuracy. To discover interpretable connections, Lei et al.~\cite{ref_22} presented an auto-weighted centralized multi-task model to combine the two kinds of connectivities for MCI study, which has achieved excellent diagnosis performance and estimated essential brain connections for further treatment. Nevertheless, the previous models adopted structural connectivity directly from the empirical methods, which may be inaccurate and ineffective for downstream feature extraction because of different manual parameter settings. Moreover, the functional connections from fMRI may be influenced by the selection of sliding time windows and the high noise.

\begin{figure*}[htbp!]
	\centering
	\includegraphics[scale=0.49]{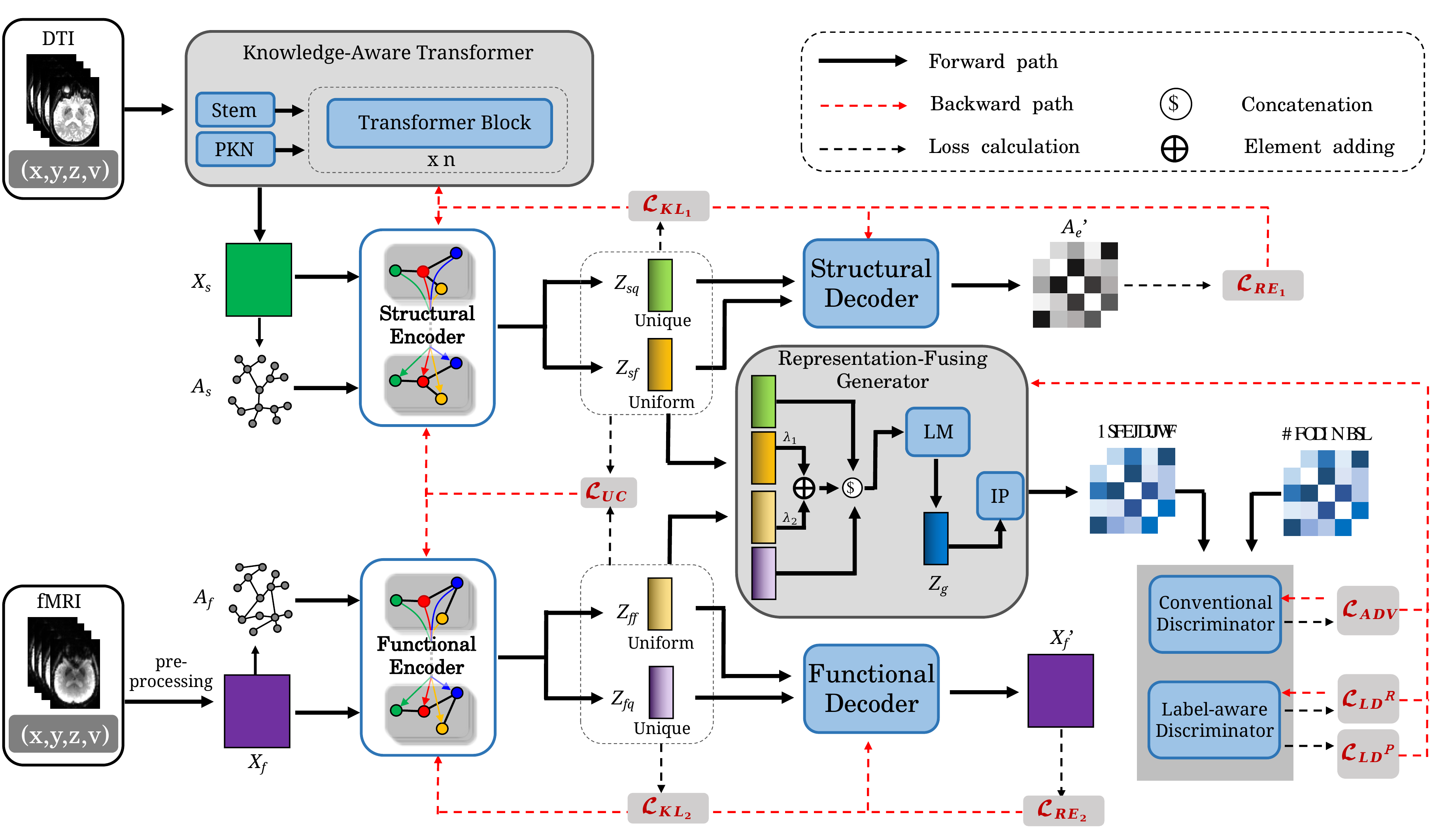}
	\caption{The overall framework of the proposed BSFL for MCI diagnosis using fMRI, DTI, and the knowledge of template-segmented brain regions. It consists of eight components: a transformer, two encoders, two decoders, a generator, and two discriminators. The output of the representation-fusing generator is the unified brain network, which is used for MCI analysis.}
	\label{fig1}
\end{figure*}

There are two strategies to learn representations from multimodal images for brain disease analysis: mixed representation learning and decomposed representation learning. The former strategy extracts latent features from unimodal data separately and fuses these features by concatenation or other specific mechanisms (i.e., averaging or weighting)~\cite{zuo2021multimodal}. Zhou et al.~\cite{ref_23} used the combination of volumetric measures calculated from T1-weighted MRI, metabolic measures generated from positron emission tomography (PET), and genetic measurements extracted from single nucleotide polymorphism (SNP) as input features to diagnose AD. Also, the work in~\cite{ref_22} merged structural connectivity features with functional connectivity features using a weighting scheme and then adopted an SVM classifier for early AD diagnosis. Decomposed representation learning via VAE has shown great potential and has become the mainstream in medical image analysis~\cite{ref_25,ref_26}. It jointly encodes each modal image into latent representations with separate meanings and combines these multimodal representations for downstream tasks. Zhang et al.~\cite{ref_27} proposed a VAE-based model to decompose multi-view brain networks from DTI and learn a unified representation, which improves MCI diagnosis performance. Similarly, Cheng et al.~\cite{ref_28} applied multimodal VAE to learn common and distinctive representations from preoperative multimodal images for glioma grading. In general, mixed representation learning may lead to common information redundancy and the degradation of the fusion effect. And the decomposed representation learning concentrates on the euclidian space for disease diagnosis, it is not suitable for brain disease analysis in terms of brain topological characteristics.

\section{Method}

\subsection{Overview}
The flowchart of BSFL is shown in Fig.~\ref{fig1}. After some preprocessing steps, given the fMRI and DTI, the proposed model learns a complicated non-linear mapping network to transform the bimodal images into brain networks for detecting abnormal brain connections at different stages of MCI. The proposed model consists of four parts: 1) knowledge-aware transformer, 2) decomposed variational graph autoencoders, 3) representation-fusing generator, and 4) dual discriminator. The last three parts are defined as the decomposition-fusion framework. First, the transformer-based network extracts structural features from DTI by incorporating location and volume information of predefined ROIs. Then, the feature space is decomposed into unique and uniform spaces for each modality by the decomposed variational graph autoencoders. Finally, the decomposed representations are fused to generate unified brain networks by the representation-fusing generator and the dual discriminator. The proposed model is featured by incorporating the following objective functions: the Kullback-Leibler (KL) loss, the reconstruction loss, the adversarial loss, the classification loss, and the uniform-unique contrastive loss. These loss functions aim to ensure decomposition thoroughness and enhance structural-functional fusion.

\begin{figure}[htbp]
	\centerline{\includegraphics[width=\columnwidth]{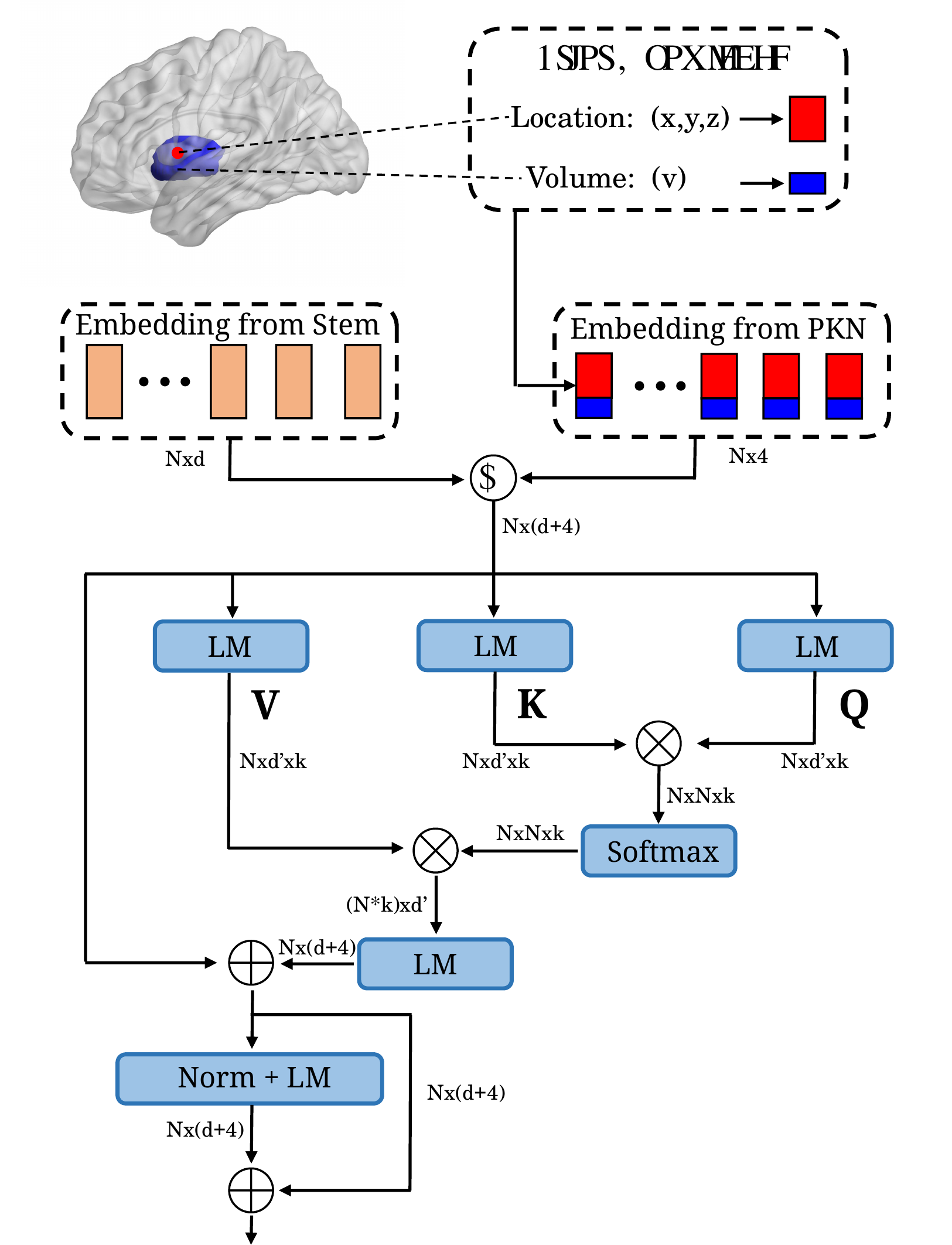}}
	\caption{The detailed structure of the transformer block in KAT. The stem embedding does not correspond to the ROIs. The output of this module is the ROI structural features.}
	\label{fig2}
\end{figure}

\subsection{Architectures}
\label{sec: architectures}
\subsubsection{Knowledge-Aware Transformer}
The transformer-based network is adopted to extract structural features for each predefined ROI from DTI in this section. The knowledge-aware transformer consists of a stem module, prior knowledge normalization (PKN) module, and a transformer block. A stem is a common form of convolution with the defined kernels and strides, which is applied to extract low-level features from each $91 \times 109 \times 91$ volume. Suppose the brain is divided into $N$ ROIs, in the proposed model, the $3 \times 3 \times 3$ convolutions~\cite{ref_29} are stacked with $2$-stride and $1$-stride at intervals, followed by a single $1\times1\times1$ convolution in the penultimate layer to match $N$-channel feature for ROIs. The filter numbers are 8, 8, 16, 16, 32, 64, and $N$ for the seven convolutional layers. In the last layer of the stem module, a one-layer linear mapping (LM) network transforms the flattened ROI feature into $d$-dimensional embedding and input to the transformer block.

Prior knowledge means the spatial location and morphology information of the predefined ROIs. According to the standard anatomical template, it can provide the central location (i.e., $\boldsymbol x$, $\boldsymbol y$, and $\boldsymbol z$) and volume (i.e., $\boldsymbol v$) information for each ROI. The PKN module normalizes the location and volume information into the range $-1\sim1$. For example, $PKN(\boldsymbol x_i)=2(\boldsymbol x_i-min(\boldsymbol x))/(max(\boldsymbol x)-min(\boldsymbol x))-1$, where, $\boldsymbol x\in\mathbb{R}^{N}$, $N$ is the number of ROIs. This formula can be applied to other prior knowledge (i.e., $\boldsymbol y$, $\boldsymbol z$, and $\boldsymbol v$).

As illustrated in Fig.~\ref{fig2}, both outputs of the stem and PKN are sent to the transformer block to learn spatial and morphological features for each ROI. One stage of the transformer block consists of prior multi-head self-attention (PMS) and feed-forward network (FFN). Here, the $F_e\in\mathbb{R}^{N \times d}$ and $F_p\in\mathbb{R}^{N \times 4}$ are denoted as the embedding of the Stem and PKN module respectively, where $d$ indicates the output dimension of the stem module. The output of each transformer block is:
\begin{equation}
	E_{out}= E_{hidden} + FFN(LM(E_{hidden}))
\end{equation}
\begin{equation}
	E_{hidden}= F_e + PMS(LM(F_e,F_p))
\end{equation}
where, $E_{hidden}\in\mathbb{R}^{N \times d}$ and $E_{out}\in\mathbb{R}^{N \times d}$. In particular, the embedding $F_e$ is projected to get value $\mathbf V$ by applying $k$ parallel linear mapping layers (i.e., heads). Also, the query $\mathbf Q$ and key $\mathbf K$ are obtained in the same way by concatenating embedding $F_e$ and $F_p$ as the input of the linear mapping layers. The dimension of the three tokens is $d'$, where, $d'=d/k$. For example, $k=1$, and the PMS can be simplified to prior single-head self-attention (PSS). It can be defined as:
\begin{equation}
	PSS(\mathbf Q, \mathbf K, \mathbf V)= Softmax(\mathbf Q \mathbf K^T / \sqrt{d'}) \mathbf V
\end{equation}
The output values of each PSS are concatenated and linearly transformed to generate the hidden result $E_{hidden}$.
Then, it is sent to FFN with one linear mapping layer and a $tanh$ activation function. Finally, the output $E_{out}$ is combined with the normalized prior knowledge $F_p$ to input the next transformer block. The output $E_{out}$ of the last transformer block is linearly mapped into $X_s$ with the dimension $d$.

\begin{figure*}[!htbp]
	\centerline{\includegraphics[scale=0.72]{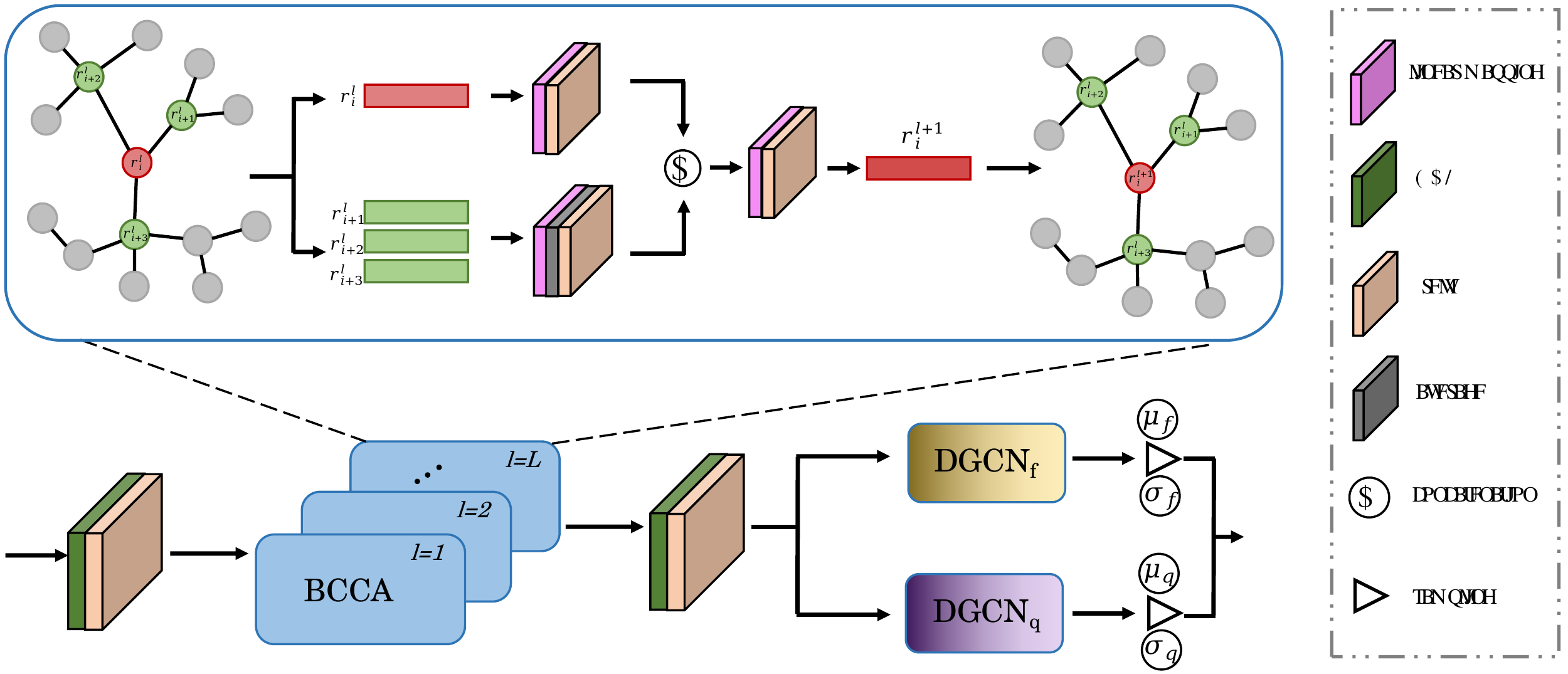}}
	\caption{The network structure of the structural encoder. It accepts both structural connectivity and structural feature, and outputs two pairs of variables: ($\mu_{f}$, $\sigma_{f}$), and ($\mu_{q}$, $\sigma_{q}$). Each pair is used to obtain the decomposed representations (i.e., uniform and unique).}
	\label{fig3}
\end{figure*}

\subsubsection{Decomposed Variational Graph Autoencoders}
After the structural features $X_s\in\mathbb{R}^{N \times d}$ and the functional features $X_f\in\mathbb{R}^{N \times d}$ extracted from the empirical method have been mined separately, decomposing the bimodal features can significantly improve the common and complementary information fusion for representation learning. Given features extracted from two modalities, this section can learn decomposed representations among modalities. There are two parts: two encoders and two decoders.

The two encoders share the same structure in Fig.~\ref{fig1}. Firstly, the structural connectivity $A_s$ is constructed by matrix inner product: $A_s = \sigma(X_s X_s^T)$. The functional connectivity $A_f$ is constructed by matrix inner product: $A_f = \sigma(X_f X_f^T)$. Then, each modal graph feature (i.e., $X_s, A_s$) is sent to encoder to get a pair of variables (i.e., $\sigma, \mu$). After that, the pair of variables are inferred to get decomposed representations. Finally, the decomposed representations are utilized to reconstruct the input features. The detailed information on the structural encoder is shown in Fig.~\ref{fig3}. The GCN is a two-layer network with $128$ and $64$ neurons. Each layer is followed by the rectified linear unit ($ReLU$) activation function. The brain connectivity central attention (BCCA) block is added to capture the global correlation between two ROIs iteratively. The feature of every ROI is updated by combing its feature and other ROIs' features. The linear mapping is one dense layer with 64 neurons. The dual graph convolutional network (DGCN) block is added to generate a pair of latent variables. It consists of two separate GCNs with one $h$-neuron layer and can output one pair of latent variables representing common and complementary information. The output of structural encoder are $\mu_{sf}$, $\sigma_{sf}$, $\mu_{sq}$, and $\sigma_{sq}$, and the output of functional encoder are $\mu_{ff}$, $\sigma_{ff}$, $\mu_{fq}$, and $\sigma_{fq}$.

To infer representations in latent space, a standard normal distribution constraint at the end of the encoder is added to get latent representations. The formula can be expressed as:
\begin{equation}\label{eq41}
	Z_{sf} = \mu_{sf} + \sigma_{sf} \odot \varepsilon_1, and\ Z_{sq} = \mu_{sq} + \sigma_{sq} \odot \varepsilon_2
\end{equation}
\begin{equation}\label{eq42}
	Z_{ff} = \mu_{ff} + \sigma_{ff} \odot \varepsilon_3, and\ Z_{fq} = \mu_{fq} + \sigma_{fq} \odot \varepsilon_4
\end{equation}
where, $\mu_{sq}$, $\sigma_{sq}$ are the mean and standard deviation matrix of the structure-specific component, while $\mu_{sf}$ and $\sigma_{sf}$ are the mean and standard deviation matrix of the uniform component in the structural encoder. The symbols in the functional encoder also have the same meaning. $\odot$ denotes element-wise product. $\varepsilon_i$ $(i \in {1,2,3,4})$ means a matrix sampled from a Gaussian distribution. $Z_{sf}$, $Z_{sq}$, $Z_{ff}$, and $Z_{fq}$ share the same size $N \times h$.

The reconstruction module can retain unimodal information and enhance the stability of the model. For structural decoder, it accepts both structure-specific representation $Z_{sq}$ and uniform representation $Z_{sf}$ and output structural adjacent matrix $A_e^{'}$ with the dimension size $N \times N$. The network is the reverse operation of the structural encoder, followed by an inner product operation and a $sigmoid$ activation function. Similarly, the functional decoder transforms the function-specific representation $Z_{fq}$ and uniform representation $Z_{ff}$ into original function time series $X^{'}_f \in\mathbb{R}^{N \times d}$ by using the inverse network structure of the functional encoder.

\subsubsection{Representation-Fusing Generator}
Since the latent representations have been decomposed into unique and uniform components, it is easy to find the best weighting parameters between different components in the fusion process. The Multi-Layer Perceptron (MLP) based generator is designed to fuse the decomposed representations to generate unified brain networks $A_p$. The generator can adaptively adjust the weight between decomposed representations, which fully reflects the common and complementary information among modalities.

In the representation-fusing generator, the uniform representations from structural-functional data are firstly added with certain weight values, then concatenated with the unique representations. The formula can be expressed as:
\begin{equation}
	Z_c = (Z_{sq} \ || \  Z_m \ || \ Z_{fq})
\end{equation}
\begin{equation}
	Z_m = \lambda_1 Z_{sf} + \lambda_2 Z_{ff}
\end{equation}
where $\lambda_1$ and $\lambda_2$ determine the relative importance of the uniform representations from the bimodal data. In the experiment, both of them are set to 0.5. $||$ means concatenation. $Z_c \in \mathbb{R}^{N \times 3h}$ is the concatenated representation. After that, a two-layer linear mapping network with $2h$ and $h$ neurons is designed to adaptively fuse the learned representations and obtain the fused representation $Z_g$:
\begin{equation}
	Z_g = LM(Z_c)
\end{equation}
Finally, the fused representation $Z_g \in \mathbb{R}^{N \times h}$ is transformed into the unified brain network through inner product (IP) operation. The predictive unified brain network $A_p \in \mathbb{R}^{N \times N}$ is expressed as:
\begin{equation}
	A_p = \sigma(Z_g Z_g^T)
\end{equation}
where, the $\sigma$ is a $sigmoid$ function.

\begin{figure}[htbp]
	\centerline{\includegraphics[width=\columnwidth]{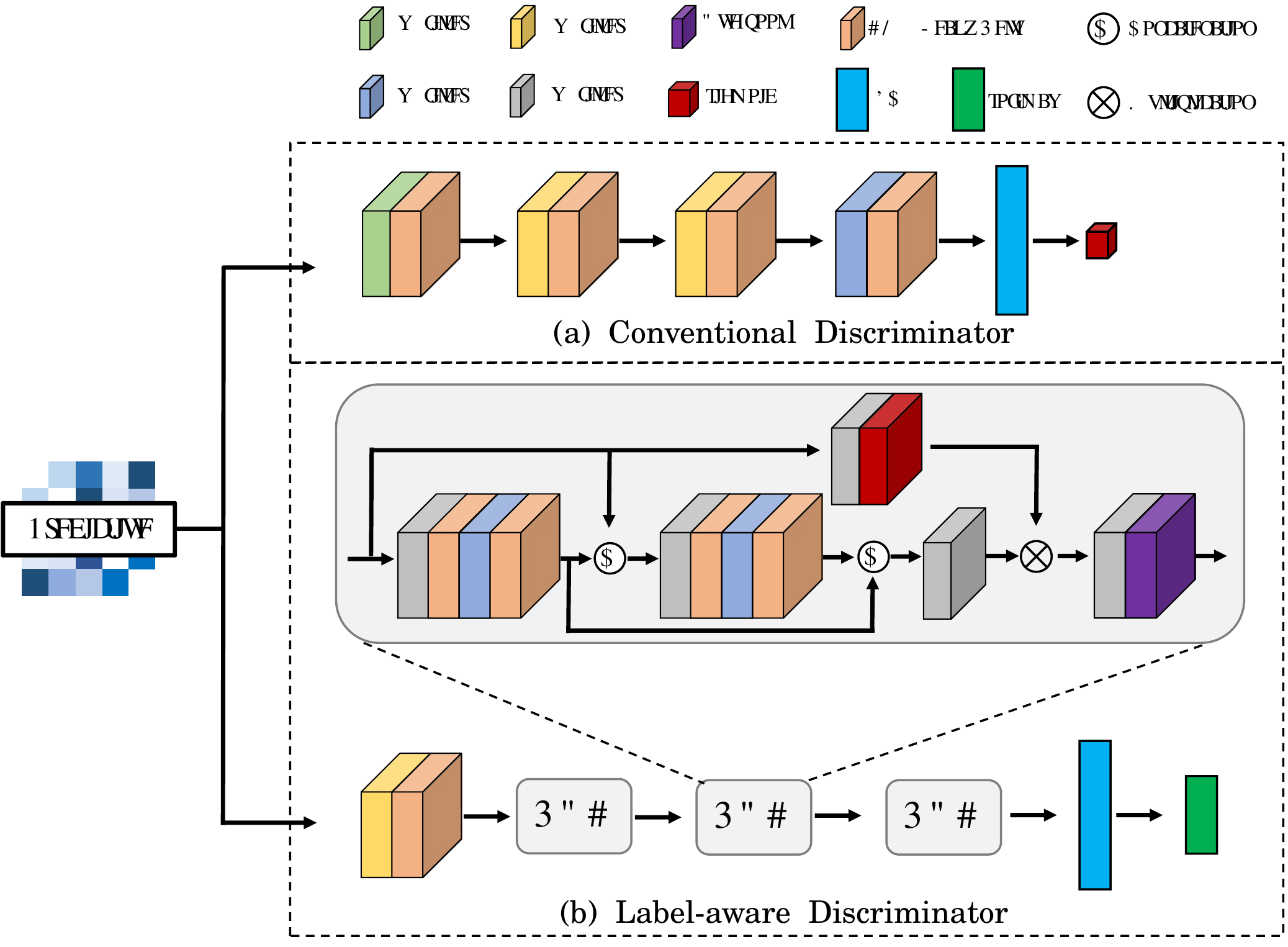}}
	\caption{The illustration of the dual discriminator in the overall framework. The label-aware discriminator is used for disease prediction.}
	\label{fig4}
\end{figure}
\subsubsection{Dual Discriminator}
As shown in Fig.~\ref{fig4}, the conventional discriminator is used to keep the output (i.e., $A_p$) consistent with the real sample (i.e., $A_b$) distribution, where the real sample is computed using the graph-based deep model (GBDM)~\cite{ref_33}. The filter numbers are 8, 16, 32, and 64, and the fully connected (FC) layer has 1024 neurons. The label-aware discriminator can classify if the input matrix is normal control or patients. It consists of one convolution layer, three residual attention blocks (RAB), and a fully-connected layer. The kernel numbers of $4 \times 4$, $3 \times 3$ and $1 \times 1$ in this discriminator are $8$, $8$, and $16$, respectively. After three RABs, a $576$-neuron FC layer and a softmax layer are added to make the feature $A_p$ discriminative.

\subsection{Hybrid loss function}
\label{sec: objective}
In this study, the KAT, encoders, decoders, generator, and discriminators are combined into the BSFL model to learn MCI-related representations and jointly trained with the following losses.
\subsubsection{KL Loss}
Assuming the latent representations obey the normal Gaussian distribution $\mathbb{N}(0,1)$, the output of encoders is defined by $E_s(X_s)$ and $E_f(X_f)$. KL divergence is adapted to constrain the output to match Gaussian distribution by introducing the reparameterization technique~\cite{ref_34}. The expression is defined below:
\begin{equation}\label{eq9}
\begin{split}
	\mathcal{L}_{KL} &= \mathcal{L}_{KL_1} + \mathcal{L}_{KL_2} \\
				   &=\mathbb{E}_{{X_s} \sim P_{DTI(X_s)}} [\emptyset _{KL} ( \mathbb{E}(E_s(X_s)) | \mathbb{N}(0,1) ) ] \\
				   &+\mathbb{E}_{{X_f} \sim P_{fMRI(X_f)}} [\emptyset _{KL} ( \mathbb{E}(E_f(X_f)) | \mathbb{N}(0,1) ) ]
\end{split}
\end{equation}
where, $\mathbb{E}$ indicates expected value, $\emptyset$ indicates KL divergence.

\subsubsection{Reconstruction Loss}
The reconstruction can stabilize the representation learning. The uniform and unique representations are combined to reconstruct the structural or functional features. Here, the $D_s$ and $D_f$ are denoted as the structural and functional decoders, respectively. $A_e$ is the empirical structural connectivity computed from the PANDA toolbox. The loss is defined as:
\begin{equation}\label{eq10}
\begin{split}
	\mathcal{L}_{RE} &= \mathcal{L}_{RE_1} + \mathcal{L}_{RE_2} \\
				  &= [\mathbb{E}_{{Z_{sf}} \sim P_{(Z_{sf})},{Z_{sq}} \sim P_{(Z_{sq})}} || A_e - D_s(Z_{sf},Z_{sq}) ||_2 \\
                  &+\mathbb{E}_{X_s \sim P_{(X_s)}} || A_e - \sigma(X_s X_s^T)) ||_2] \\
			      &+\mathbb{E}_{{Z_{ff}} \sim P_{(Z_{ff})},{Z_{fq}} \sim P_{(Z_{fq})}} || X_f - D_f(Z_{ff},Z_{fq}) ||_2
\end{split}
\end{equation}

\subsubsection{Adversarial Loss}
The output of the representation-fusing generator is the unified brain network, which is defined as $A_p$. Here, $A_p = G(Z_{sf},Z_{sq},Z_{ff},Z_{fq})$. The benchmark brain network is denoted as $A_b$, computed from the GBDM method by deeply fusing fMRI and DTI. The conventional discriminator is represented as $D_c$. The adversarial loss can be written as:
\begin{equation}\label{eq11}
	\mathcal{L}_  {D} = \mathbb{E}_{{A_p} \sim P_{A_p}}[(D_c(A_p))^2] + \mathbb{E}_{{A_b} \sim P_{A_b}}[(D_c(A_b)-1)^2]
\end{equation}
\begin{equation}\label{eq12}
	\mathcal{L}_  {G} = \mathbb{E}_{{A_p} \sim P_{A_p}}[(D_c(A_p)-1)^2]
\end{equation}
\begin{equation}\label{eq13}
\mathcal{L}_  {ADV} = \mathcal{L}_  {G} + 0.1 \mathcal{L}_  {D}
\end{equation}


\subsubsection{Classification Loss}
To discriminate the unified brain network, a label-aware discriminator is defined as $D_l$ to classify if $A_p$ and $A_r$ are normal control or patient. The formula is defined as:
\begin{equation}\label{eq14}
\begin{split}
	\mathcal{L}_  {CL} &= \mathcal{L}^R_{LD} + \mathcal{L}^P_{LD} \\
				    &=\mathbb{E}_{A_b \sim P_{(A_b)}}[-I \cdot log(D_l(A_b))] \\
				    &\ \ \ +\mathbb{E}_{A_p \sim P_{(A_p)}}[-I \cdot log(D_l(A_p))]
\end{split}
\end{equation}
where, $I$ is the truth label.

\subsubsection{Uniform-unique Contrastive Loss}
To constrain the learned decomposed representations, a uniform-unique contrastive (UC) loss function is applied to constrain the distance between them. The expression is defined as:
\begin{equation}\label{eq15}
\begin{split}
	\mathcal{L}_  {UC} &= \frac{1}{4} ( \mathcal{L}_{UC_1} + \mathcal{L}_{UC_2} ) + \frac{1}{2} \mathcal{L}_{UC_3} \\
	&=\frac{1}{4} \mathbb{E}_{Z_{sq},Z_{sf}}[ max(margin - || (Z_{sq} - Z_{sf}) ||_2,0)] \\
	&\ \ \ +\frac{1}{4} \mathbb{E}_{Z_{fq},Z_{ff}}[ max(margin - || (Z_{fq} - Z_{ff}) ||_2,0)] \\
    &\ \ \ +\frac{1}{2} \mathbb{E}_{Z_{sf},Z_{ff}}(|| Z_{sf} - Z_{ff} ||_2)
\end{split}
\end{equation}
here, $margin$ indicates the threshold with default value 1.

The detailed training steps of the proposed BSFL are described in Algorithm~\ref{algorithm1} for reference.

\begin{algorithm}[htb]
	\caption{The optimization procedure for the BSFL}
	\label{algorithm1}
	\begin{algorithmic}[1]
		\Require
		each mode's features $DTI$ and $X_f$, empirical struc-\newline
        \hspace*{1.2em}  tural connectivity $A$, real unified brain network $A_r$,\newline
		\hspace*{1.2em}  maximal iterative number $maxIter$, training step \newline
		\hspace*{1.2em}  parameter $t$, model parameters $\Theta$, hyper-parameters \newline
		\hspace*{1.2em}  $\lambda _1, \lambda _2$ (set both as 0.5), prior knowledge $x,y,z,v$.
		\Ensure
		predictive unified brain network $A_p$, model parame-\newline
        \hspace*{1.2em}  ters $\Theta$;
		\State initialization: $\Theta$, $t$, $maxIter$.
		\Repeat
		\State $t \gets t+1$
		\State compute structural features based on the KAT module: \newline
		\hspace*{1.2em} $X_s = KAT(DTI,x,y,z,v)$;
		\State compute the structural uniform and unique variables \newline
        \hspace*{1.2em}  based on the encoder $E_s$: \newline
        \hspace*{1.2em}  ($\mu_{sf}$, $\sigma_{sf}$)=$Uniform(E_s(X_s))$, \newline
        \hspace*{1.2em}  ($\mu_{sq}$, $\sigma_{sq}$)=$Unique(E_s(X_s))$;
		\State compute the functional uniform and unique variables:\newline
        \hspace*{1.2em}  based on the encoder $E_f$: \newline
        \hspace*{1.2em}  ($\mu_{ff}$, $\sigma_{ff}$)=$Uniform(E_f(X_f))$, \newline
        \hspace*{1.2em}  ($\mu_{fq}$, $\sigma_{fq}$)=$Unique(E_f(X_f))$;
        \State  compute the uniform and unique representations based \newline
        \hspace*{1.2em}  on the above variables and the random noise $\varepsilon$ sampled \newline
        \hspace*{1.2em}  from a standard normal distribution $\mathbb{N}(0,1)$:\newline
        \hspace*{1.2em}  $Z_{sf} = \mu_{sf} + \sigma_{sf} \odot \varepsilon_1, \ Z_{sq} = \mu_{sq} + \sigma_{sq} \odot \varepsilon_2$,\newline
        \hspace*{1.2em}  $Z_{ff} = \mu_{ff} + \sigma_{ff} \odot \varepsilon_3, \ Z_{fq} = \mu_{fq} + \sigma_{fq} \odot \varepsilon_3$;
		\State compute the unified brain network $A_p$ based on \newline
        \hspace*{1.2em} the representation-fusing generator $G$:\newline
        \hspace*{1.2em} $A_p = G(Z_{sf},Z_{sq},Z_{ff},Z_{fq})$;
		\State reconstruct the input features $A'$ and $X^{'}_f$ with the\newline
		\hspace*{1.2em} decoders $D_s$ and $D_f$: \newline
        \hspace*{1.2em} $A_e'=D_s(Z_{sf},Z_{sq})$, $X^{'}_f=D_f(Z_{ff},Z_{fq})$;
        \State Update the $\Theta$ in conventional discriminator $D_c$ by\newline
        \hspace*{1.2em} back propagating the gradient $\nabla_{\Theta} L_{D}^{t}$;
		\State add the KL loss Eq.(~\ref{eq9}), the reconstruct loss Eq.(~\ref{eq10}),\newline
		\hspace*{1.2em} the generator loss Eq. (~\ref{eq12}), the classification loss\newline
		\hspace*{1.2em} Eq.(~\ref{eq14}) and the uniform-unique contrastive loss \newline
		\hspace*{1.2em} Eq.(~\ref{eq15}) to the loss $L_{merge}^{t}$;
		\State calculate the gradient loss $\nabla_{\Theta} L_{merge}^{t}$;
		\State replace $A_p$ with $A^t_p$ and update the $\Theta$ in encoders, de-\newline
        \hspace*{1.2em} coders, generators and label-aware discriminator by \newline
		\hspace*{1.2em} taking adaptive gradient steps.
		\Until{$t>maxIter$}
	\end{algorithmic}
\end{algorithm}


\section{Experiments}
\label{sec: Experiments1}
\subsection{Data description and preprocessing}
There are four stages associated with cognitive disease degeneration in this work, including normal control (NC), significant memory concern (SMC), early mild cognitive impairment (EMCI), and late mild cognitive impairment (LMCI). About 324 subjects with both DTI and fMRI were selected from the Alzheimer's Disease Neuroimaging Initiative (ADNI)\footnote{http://adni.loni.usc.edu/} database for the proposed model. Table~\ref{Table0} lists the detailed information about the selected subjects. Note that SMC is the transitional stage from NC to EMCI. EMCI and LMCI are subtypes of MCI.

\begin{table}[htbp]
    \renewcommand\arraystretch{1.2}
	\caption{Summary of the subject information in the experiment.}\label{Table0}
	\begin{center}
		\begin{tabular}{ccccc}
			\toprule
			\textbf{Group} & \textbf{NC(82)} & \textbf{SMC(82)} & \textbf{EMCI(82)} & \textbf{LMCI(76)}\\
			\midrule
			Male/Female & 39M/43F & 35M/47F & 40M/42F & 43M/33F\\
			Age(mean$\pm$SD) & 74.2/8.1 & 76.1/5.4 & 75.9/7.5 & 75.8/6.4\\
			\bottomrule
		\end{tabular}
	\end{center}
\end{table}

The DTI scanning resolution ranges from $0.9mm$ to $2.7mm$ in X and Y directions with a slice thickness of $2.0mm$. The gradient directions are in the range $6\sim126$. The parameters TR (Time of Repetition) and TE (Time of Echo) are in the range of $3.4s \sim 17.5s$ and $56ms \sim 105ms$, respectively. PANDA toolbox~\cite{ref_35} is used to perform sample preprocessing on DTI to get a fraction anisotropy (FA) image with the dimensional size $91\times109\times91$.

The fMRI scanned by a 3T MRI equipment has different slice thicknesses in the range of $2.5mm\sim3.4mm$. The image resolution ranges from $2.5mm$ to $3.75mm$ in both plane dimensions. The TR value is between $0.607s$ to $3.0s$, and the TE value ranges from $30ms$ to $32ms$. The duration of scanning data is 10 minutes. In the preprocessing stage, the GRETNA toolbox~\cite{ref_37} is utilized to acquire functional time series based on the Automated Anatomical Labelling (AAL) atlas~\cite{ref_38}. The preprocessing steps include correcting head-motion artifacts, spatial normalization, smoothing, removing linearized drift, and band-pass filtering. Finally, the 90 non-overlapping ROIs time series are obtained by normalizing them into the same TR value. The output of this procedure is the input feature $X_f$ of the proposed model with the dimension size $90\times187$.

\subsection{Experimental settings}
In this experiment, three binary classification tasks are performed, i.e., (1) SMC vs. NC, (2) EMCI vs. NC, and (3) LMCI vs. NC. The experiments are conducted using 10-fold cross-validation to ensure the results are stable. Also, the proposed model is compared with other related methods to demonstrate the proposed model's superiority. There are three methods in the comparison: (1) the empirical method that derives the structural connectivity (SC) and static functional connectivity (FC) from the commonly used software toolboxes (i.e., PANDA, and GRETNA) and then averages them to obtain empirical brain networks; (2) the GBDM method that deeply fuses the SC and functional time series and then generates benchmark brain networks; (3) our method that transforms the DTI and functional time series to unified brain networks. The results of each methods are sent to the same classifier (i.e., SVM~\cite{ref_39}, DNN~\cite{ref_43}, and GCN~\cite{ref_44}) to compare their prediction performance.

The model parameter settings in the experiments are: $N=90, d=187, q=64, h=32, n=3$. TensorFlow1\footnote{http://www.tensorflow.org/}  is utilized to implement on an NVIDIA TITAN RTX2080 GPU device to train a convergent model for 10 hours with about 600 epochs. The initial learning rate of the transformer, encoders, and decoders is $10^{-3}$ and will decrease to $10^{-4}$ after 200 epochs. The learning rate of the generator and the conventional discriminator are set to 0.0001 and 0.0004 respectively. For the label-aware discriminator, the learning rate is set to 0.0001. The dropout ratio in both generator and the dual discriminator is set at 0.5. The Adam is adopted to optimize the training process with batch size 16.

\subsection{Unified brain network analysis}

\begin{figure}[htbp]
	\centerline{\includegraphics[width=\columnwidth]{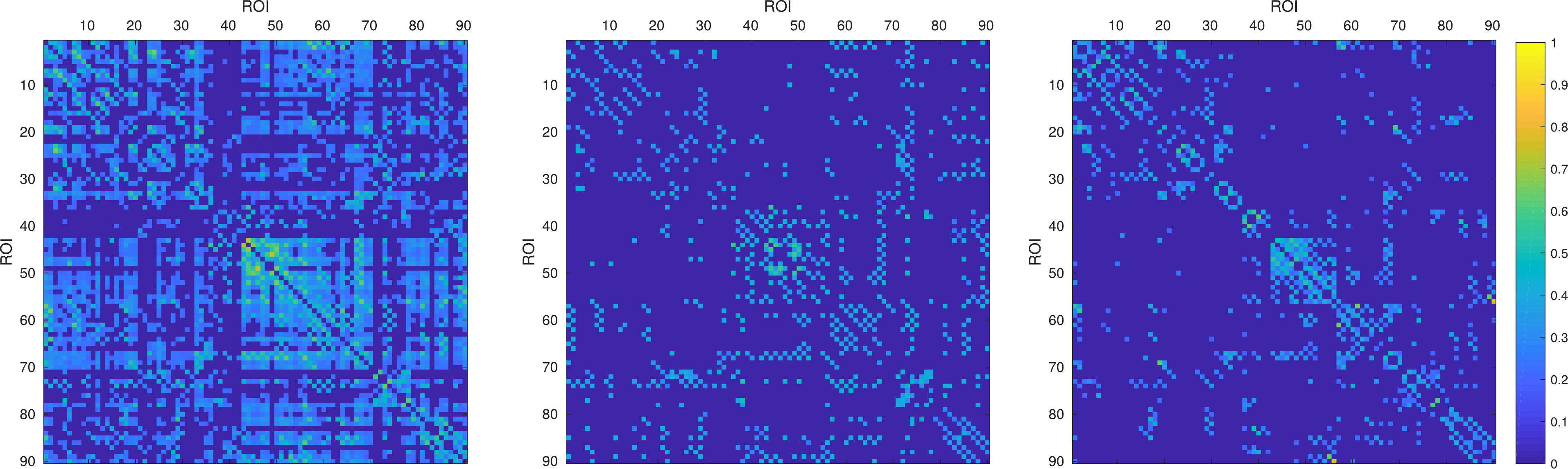}}
	\caption{An example of the generated brain networks using three methods: the empirical method (left), the GBDM method (middle), and the proposed method (right).}
	\label{fig8}
\end{figure}

The proposed model aims to generate unified brain networks for disease analysis. This section analyzes the generated brain networks in terms of prediction tasks. To compare the performance of our method with other related methods, three binary prediction tasks are conducted to calculate the mean values of evaluation indicators (i.e., ACC, SEN, SPE, and AUC) using a cross-validation strategy. Three different classifiers are adopted to evaluate the prediction performance of the generated brain networks. Fig.~\ref{fig8} shows a qualitative example of three brain networks using different methods. The quantitative prediction results are displayed in Table~\ref{Table1}.  The results show that our method achieves the best prediction performance among the three methods. It may indicate that the proposed model can make full use of the common and complementary information from the structural-functional data and thus make the fusion more effective.

\begin{table*}[htbp]
	\centering
	\renewcommand\arraystretch{1.2}
	\setlength{\abovecaptionskip}{0pt}%
	\setlength{\belowcaptionskip}{10pt}%
	\caption{Mean prediction results of the generated brain networks by different classifiers.(\%)}\label{Table1}
	\begin{tabular}{c|c|cccc|cccc|cccc}
		\hline
		\multirow{2}{*}{\textbf{Classifier}}   & \multirow{2}{*}{\textbf{Brain networks generated by}} & \multicolumn{4}{c|}{\textbf{SMC vs. NC}} & \multicolumn{4}{c|}{\textbf{EMCI vs. NC}} & \multicolumn{4}{c}{\textbf{LMCI vs. NC}} \\ \cline{3-14}
		&                         & ACC    & SEN   & SPE  & AUC   & ACC    & SEN    & SPE   & AUC   & ACC    & SEN    & SPE   & AUC   \\ \hline
		\multirow{3}{*}{SVM}
		& Empirical   & 76.82 & 85.36 & 68.29 & 77.84 & 78.05 & 73.17 & 82.93 & 75.04 & 82.27 & 82.89 & 81.70 & 83.27 \\
		& GBDM        & 77.43 & 85.36 & 69.51 & 78.84 & 79.88 & 73.17 & 86.59 & 79.95 & 85.44 & 84.21 & 85.59 & 84.53 \\
		& Ours       & {\bfseries 79.88} & {\bfseries 89.02} & {\bfseries 70.73} & {\bfseries 87.09} & {\bfseries 84.16} & {\bfseries 74.39} & {\bfseries 93.90} & {\bfseries 95.05} & {\bfseries 87.97} & {\bfseries 88.16} & {\bfseries 87.80} & {\bfseries 95.31} \\  \hline
		\multirow{3}{*}{DNN~\cite{ref_43}}
		& Empirical   & 76.83 & 81.71 & 71.95 & 77.54 & 81.70 & 85.36 & 78.04 & 86.85 & 84.17 & 78.94 & 89.02 & 82.50 \\
		& GBDM        & 78.66 & 81.71 & 75.61 & 78.42 & 83.53 & 86.58 & 80.48 & 88.17 & 86.07 & 82.89 & 89.02 & 86.26 \\
		& Ours       & {\bfseries 82.92} & {\bfseries 84.15} & {\bfseries 81.71} & {\bfseries 86.92} & {\bfseries 89.02} & {\bfseries 87.80} & {\bfseries 90.24} & {\bfseries 96.06} & {\bfseries 91.77} & {\bfseries 88.16} & {\bfseries 95.12} & {\bfseries 94.74} \\ \hline
		\multirow{3}{*}{GCN~\cite{ref_44}}
		& Empirical   & 79.26 & 87.80 & 70.73 & 78.77 & 84.75 & 89.02 & 80.48 & 86.91 & 88.60 & 85.52 & 91.46 & 88.17 \\
		& GBDM        & 81.70 & 89.02 & 74.39 & 81.81 & 86.58 & 89.02 & 84.14 & 89.75 & 91.13 & 86.84 & 95.12 & 89.68 \\
		& Ours        & {\bfseries 85.97} & {\bfseries 96.34} & {\bfseries 75.61} & {\bfseries 93.88} & {\bfseries 90.85} & {\bfseries 93.90} & {\bfseries 87.80} & {\bfseries 97.85} & {\bfseries 94.30} & {\bfseries 93.42} &  95.12 & {\bfseries 98.12} \\ \hline
	\end{tabular}
\end{table*}

\begin{figure}[htbp]
	\centerline{\includegraphics[width=\columnwidth]{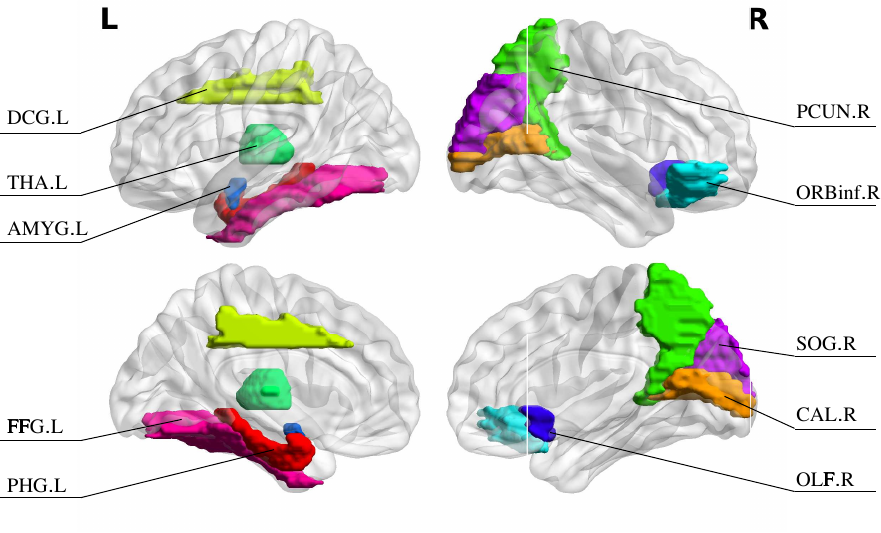}}
	\caption{The spatial distribution of top 10 related ROIs for SMC vs. NC.}
	\label{fig9_1}
\end{figure}

\begin{figure}[htbp]
	\centerline{\includegraphics[width=\columnwidth]{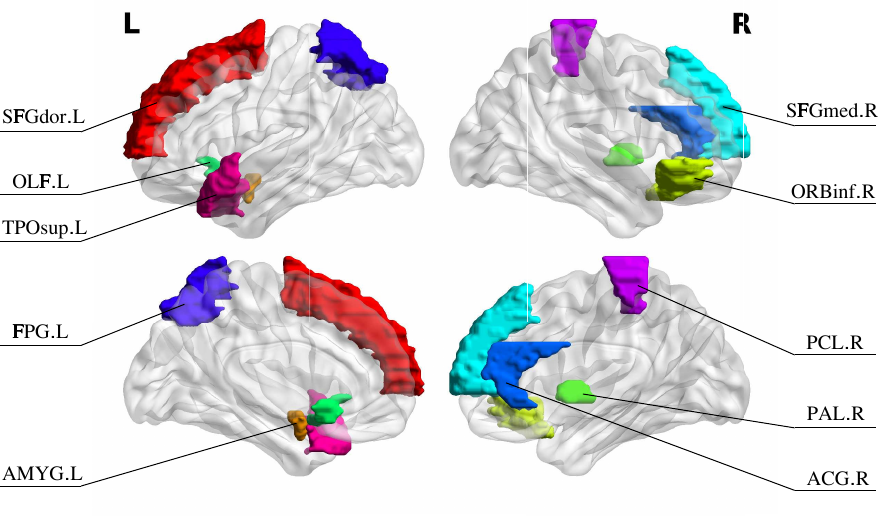}}
	\caption{The spatial distribution of top 10 related ROIs for EMCI vs. NC.}
	\label{fig9_2}
\end{figure}

\begin{figure}[htbp]
	\centerline{\includegraphics[width=\columnwidth]{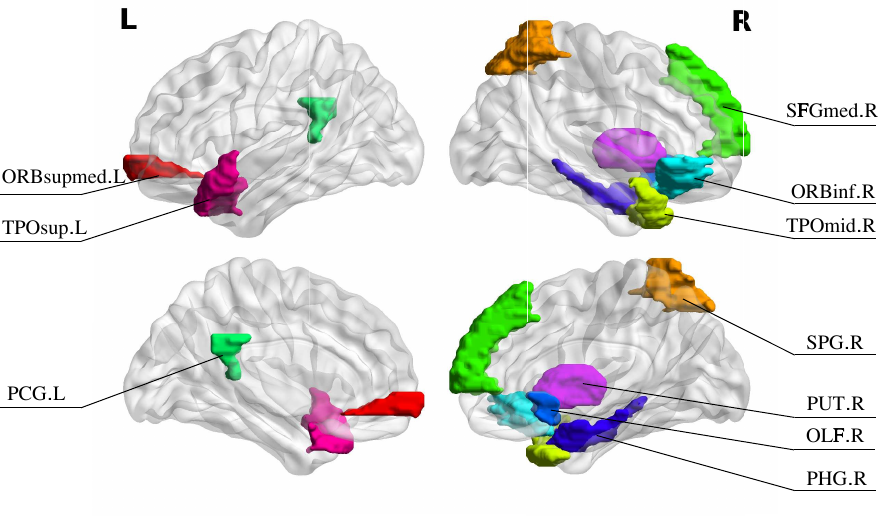}}
	\caption{The spatial distribution of top 10 related ROIs for LMCI vs. NC.}
	\label{fig9_3}
\end{figure}

To analyze the effect of different brain regions on the prediction tasks, the method that shields one brain region and calculates an importance score for each ROI is used for disease analysis. After sorting the important scores of all ROIs, the top 10 disease-related ROIs are displayed in Fig.~\ref{fig9_1}, Fig.~\ref{fig9_2} and Fig.~\ref{fig9_3}. Specifically, the top 10 related ROIs are PHG.L, CAL.R, DCG.L, PCUN.R, THA.L, ORBinf.R, AMYG.L, OLF.R, SOG.R, and FFG.L in the SMC vs. NC prediction task. For EMCI vs. NC, the ten important ROIs are in the frontal lobe (SFGdor.L, ORBinf.R, OLF.L, SFGmed.R), temporal lobe (AMYG.L, TPOsup.L), parietal lobe (SPG.L, PCL.R), subcortical area (PAL.R). The relevant brain ROIs for LMCI vs. NC is in the frontal lobe (ORBsupmed.L, SFGmed.R, ORBinf.R, OLF.R), parietal lobe (SPG.R, PCG.L), temporal lobe (TPOmid.R, PHG.R, TPOsup.L), subcortical area (PUT.R).

The prediction performance is calculated by meaning the values of accuracy (ACC), sensitivity (SEN), specificity (SPE), F1-score, and the area under the receiver operating characteristic curve (AUC). The AUC is used to evaluate the classifier's overall performance ($0\le AUC\le 1$).

\subsection{Prediction of abnormal brain connections}
\begin{figure}[htbp]
	\centerline{\includegraphics[width=\columnwidth]{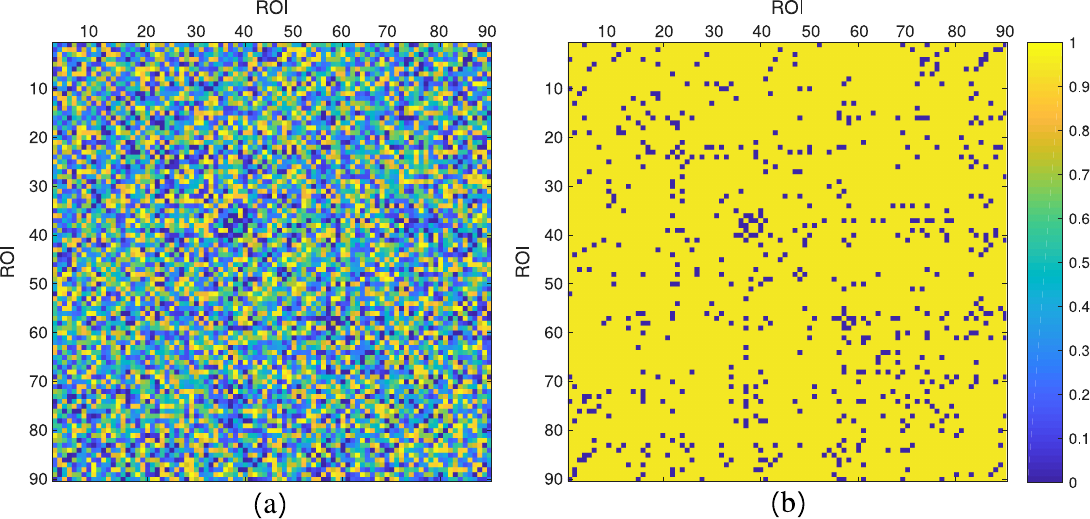}}
	\caption{The difference of brain connections between LMCI and NC groups by the proposed model. (a) The $p$-values on the connection between all ROIs. (b) The significant connections are denoted blue with $p$-value $< 0.05$.}
	\label{fig10}
\end{figure}

\begin{figure}[htbp]
	\centerline{\includegraphics[width=\columnwidth]{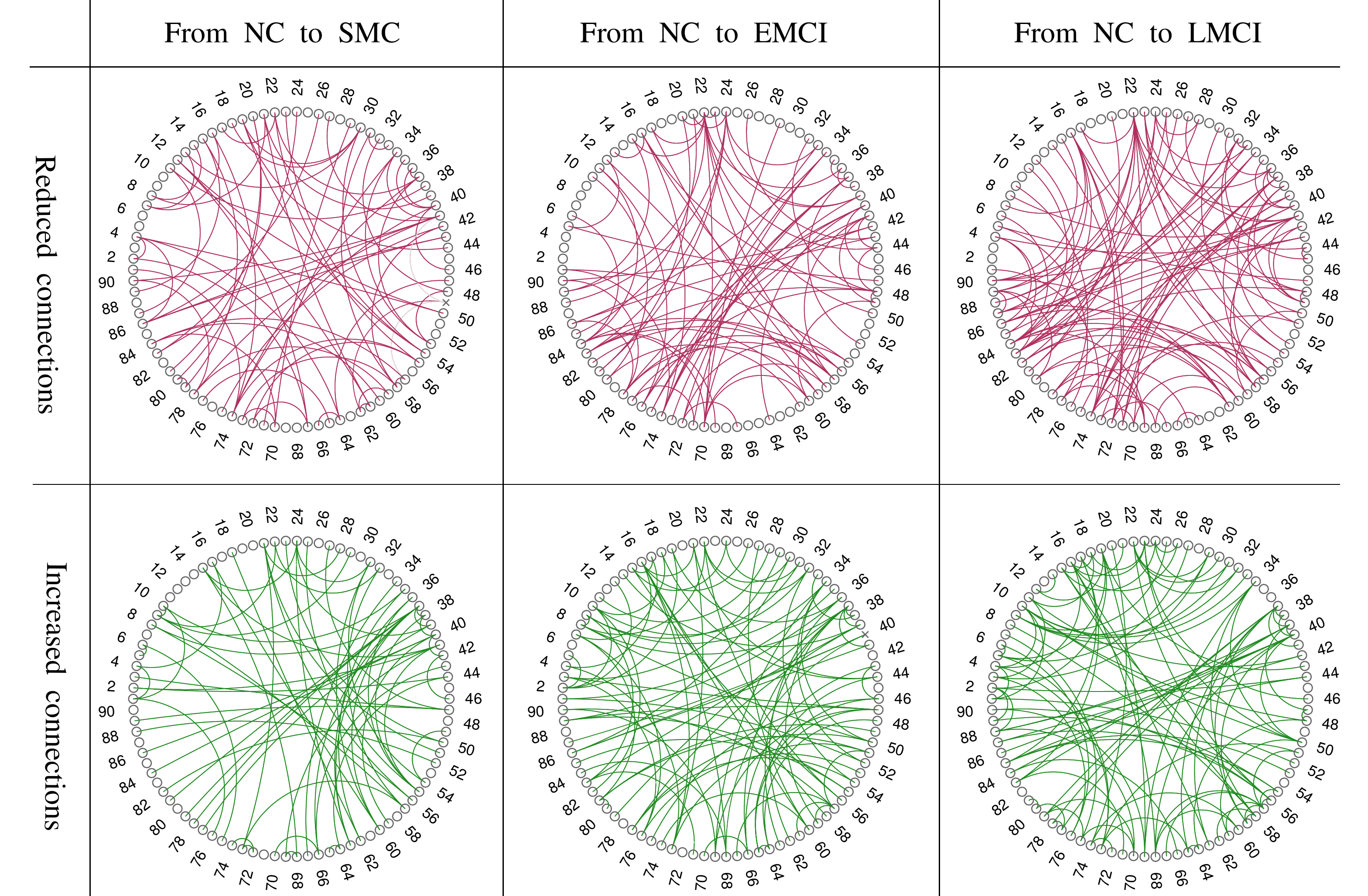}}
	\caption{Visualization of altered connections at different stages of MCI.}
	\label{fig11}
\end{figure}

In this section, predicting abnormal brain connections is studied during the cognitive disease progression. After the model has been trained to converge, it can output the unified brain network (UBN) for each subject with bimodal images (i.e., fMRI and DTI). Based on the standard two-sample t-test method, it can evaluate the significant brain connections between two groups (e.g., LMCI vs. NC) by setting the $p$-value threshold. Fig.~\ref{fig10} shows an example of $p$-values on brain connections between LMCI and NC groups by setting the threshold lower than 0.05. For ease of visualization, the significant brain connections are denoted as blue color. The most densely connected ROIs are mostly overlapped with the results of the above section. Fig.~\ref{fig11} displays the circular graph of significant brain connections at different disease stages. Compared with the normal controls, the number of reduced connections is 154, 162, and 218, while the number of increased connections is 138, 192, and 233 for SMC, EMCI, and LMCI, respectively.

\begin{figure}[htbp]
	\centerline{\includegraphics[width=\columnwidth]{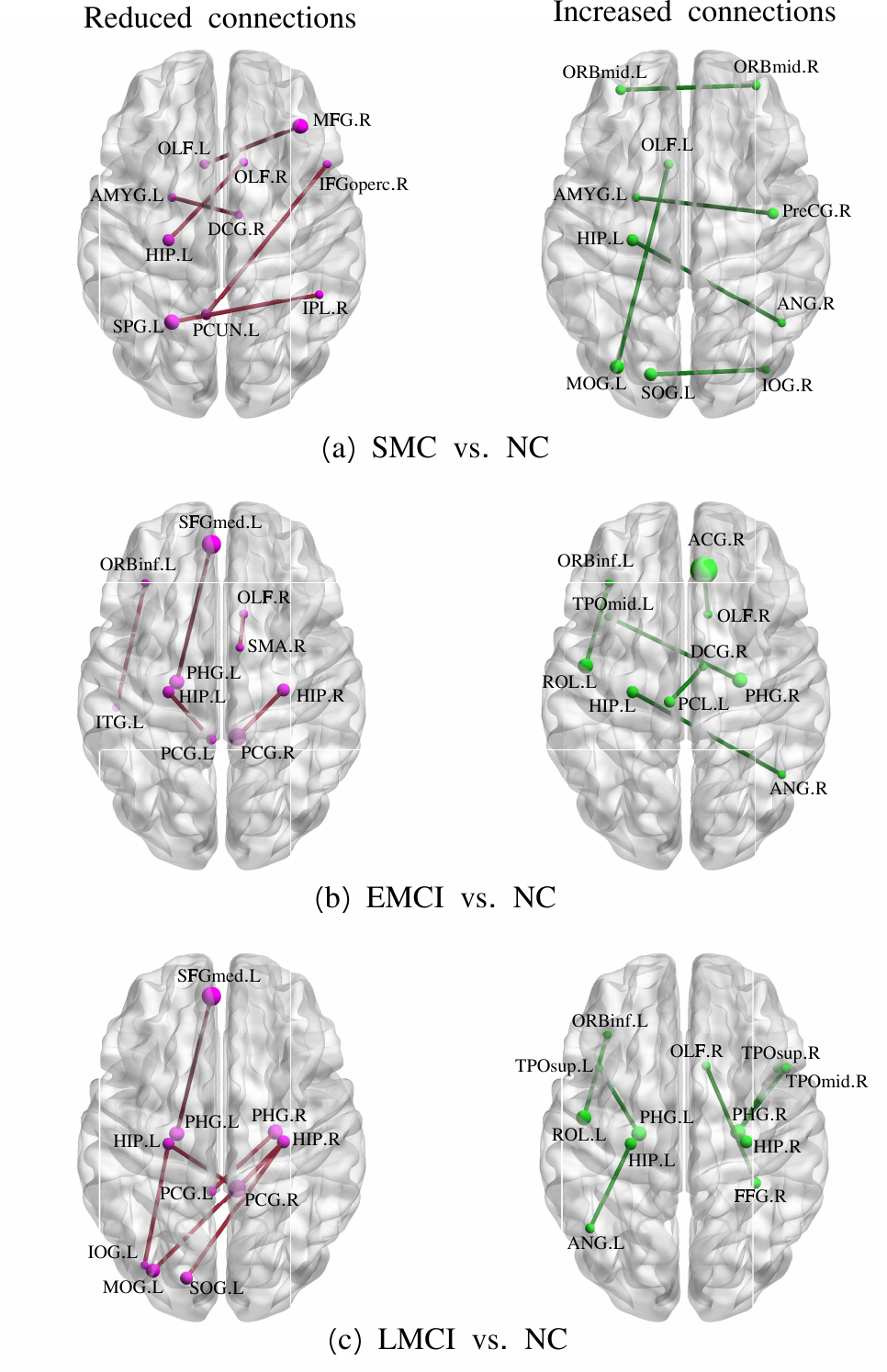}}
	\caption{The spatial visualization of important connections at different MCI stages.}
	\label{fig12}
\end{figure}

\begin{figure*}[htbp]
	\centering
	\includegraphics[scale=1.15]{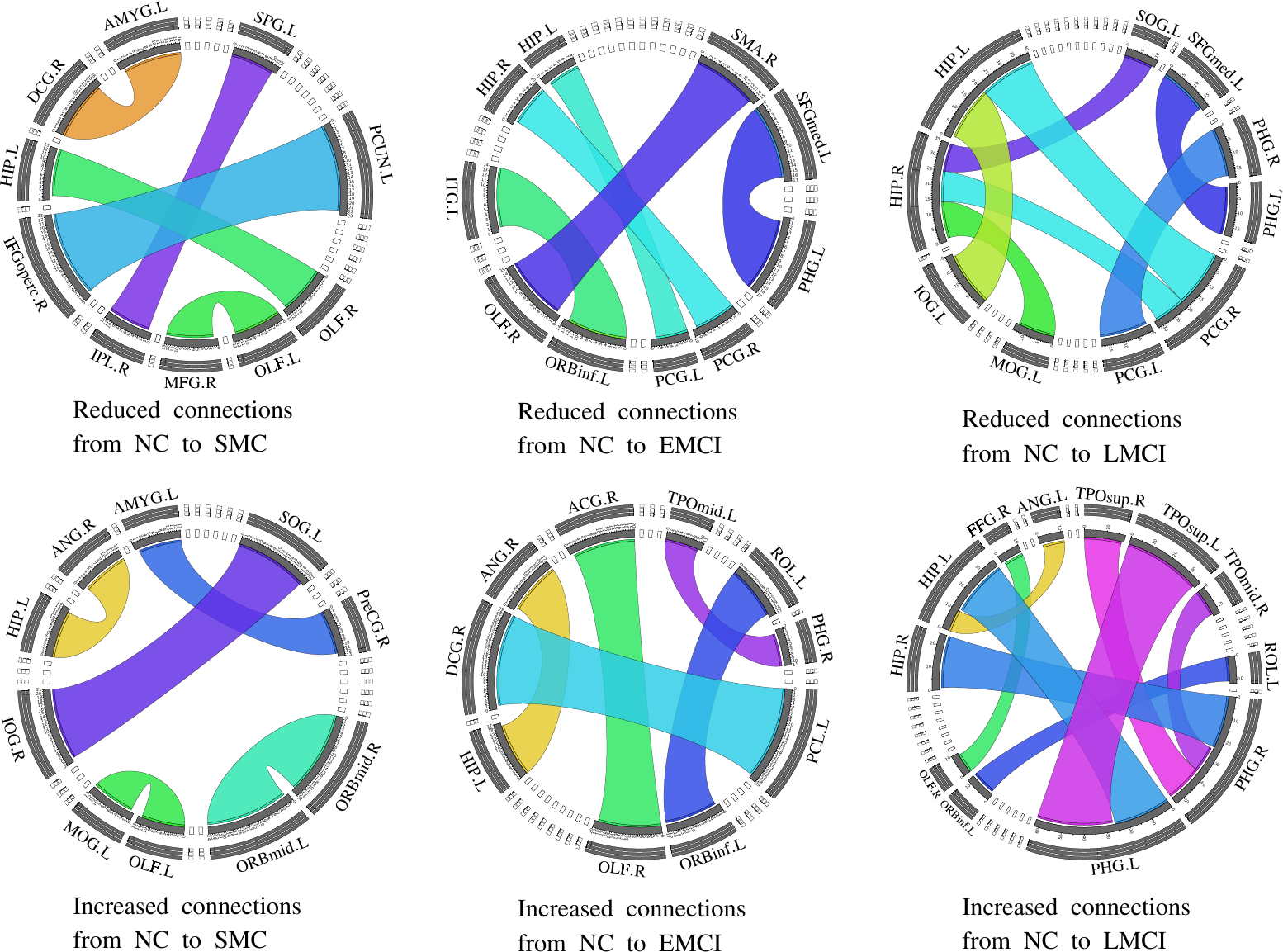}
	\caption{Visualization of important connections between two groups (i.e., SMC vs. NC, EMCI vs. NC, and LMCI vs. NC) with $p$-value $ < 0.001$. }
	\label{fig13}
\end{figure*}

To analyze the important abnormal connections, significant connections with a $p$-value lower than 0.001 are selected. The spatial location of the important abnormal connections is shown in Fig.~\ref{fig12} using the BrainNet Viewer toolbox~\cite{ref_46}. The red color means reduced connections, and the green color indicates increased connections. The ROI size defined in the AAL template means the relative volume. Fig.~\ref{fig13} shows 10, 10, and 15 important abnormal connections for SMC vs. NC, EMCI vs. NC, and LMCI vs. NC, respectively. Colors have no special meaning. The connection strength is computed by averaging UBNs for each group (i.e., NC, SMC, EMCI, and LMCI) and then subtracting the mean UBN of the NC group from that of the patient's group. For the convenience of drawing by circos table viewer~\cite{ref_45}, the connection strength is expanded by a factor of 100 and rounded up.

\subsection{Ablation Study}
\begin{figure}[htbp]
	\centerline{\includegraphics[width=\columnwidth]{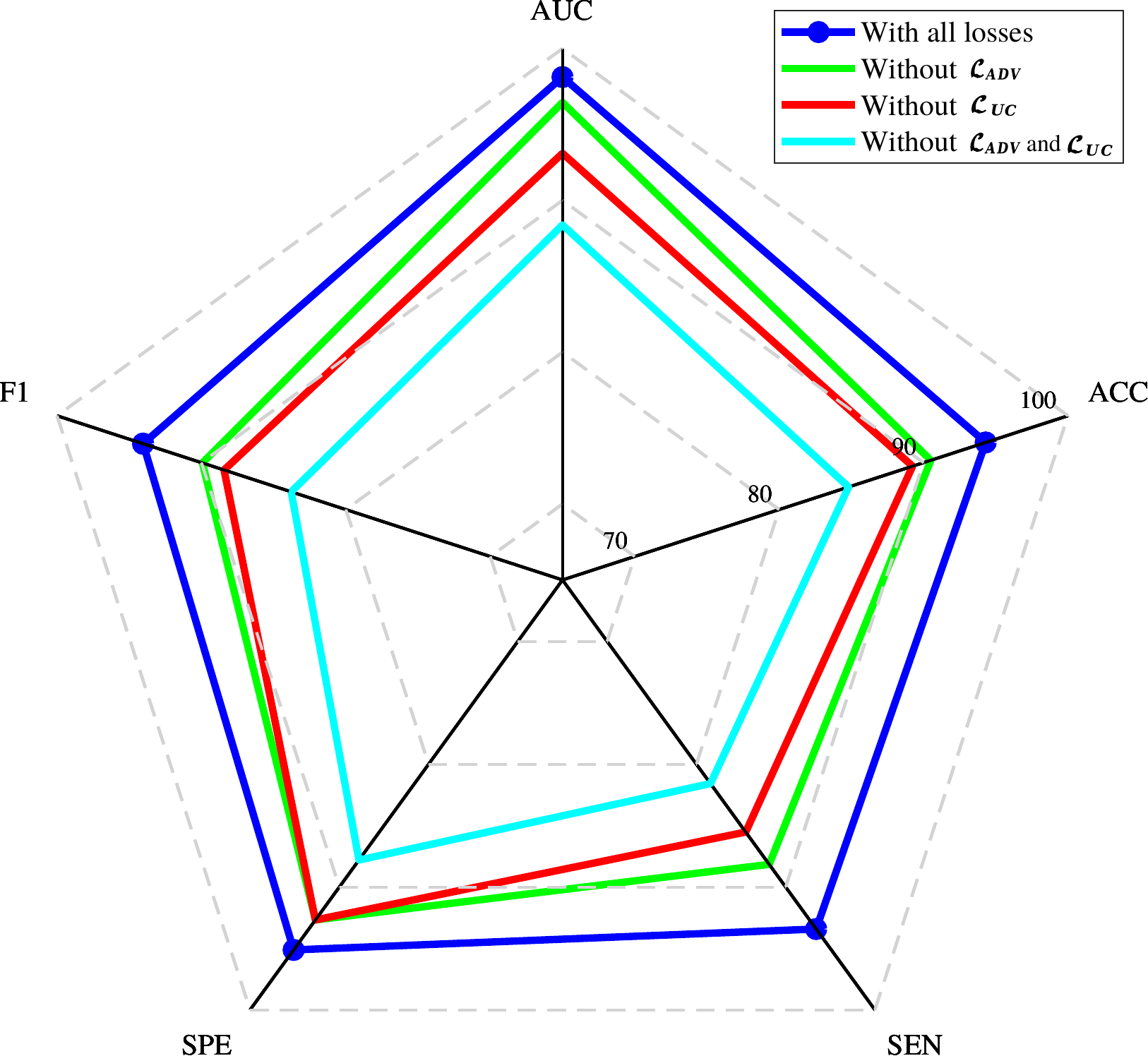}}
	\caption{Comparison of different loss functions on the prediction performance of LMCI vs. NC.}
	\label{fig14}
\end{figure}

The decomposition-fusion framework in the proposed model is essential for constructing unified brain networks. To investigate the effectiveness of the decomposed and fused modules, the uniform-unique contrastive and adversarial losses are explored in the prediction performance. Three prediction experiments are conducted in this section:  1) remove the uniform-unique contrastive loss $\mathcal{L}_{UC}$, which means the unique and uniform representations are mixed; 2) remove the adversarial loss $\mathcal{L}_{ADV}$; 3) remove both losses in the proposed model. The effects of removing different loss functions on prediction performance are shown in Fig.~\ref{fig14}. The results demonstrate that either uniform-unique contrastive loss or adversarial loss can affect the prediction performance to some extent. Both loss functions can effectively improve the proposed model's performance in terms of ACC, SEN, SPE, F1, and AUC.

\section{Discussion}
\subsection{Effectiveness of the Generator and the Decoders}
\begin{figure}[htbp]
	\centerline{\includegraphics[width=\columnwidth]{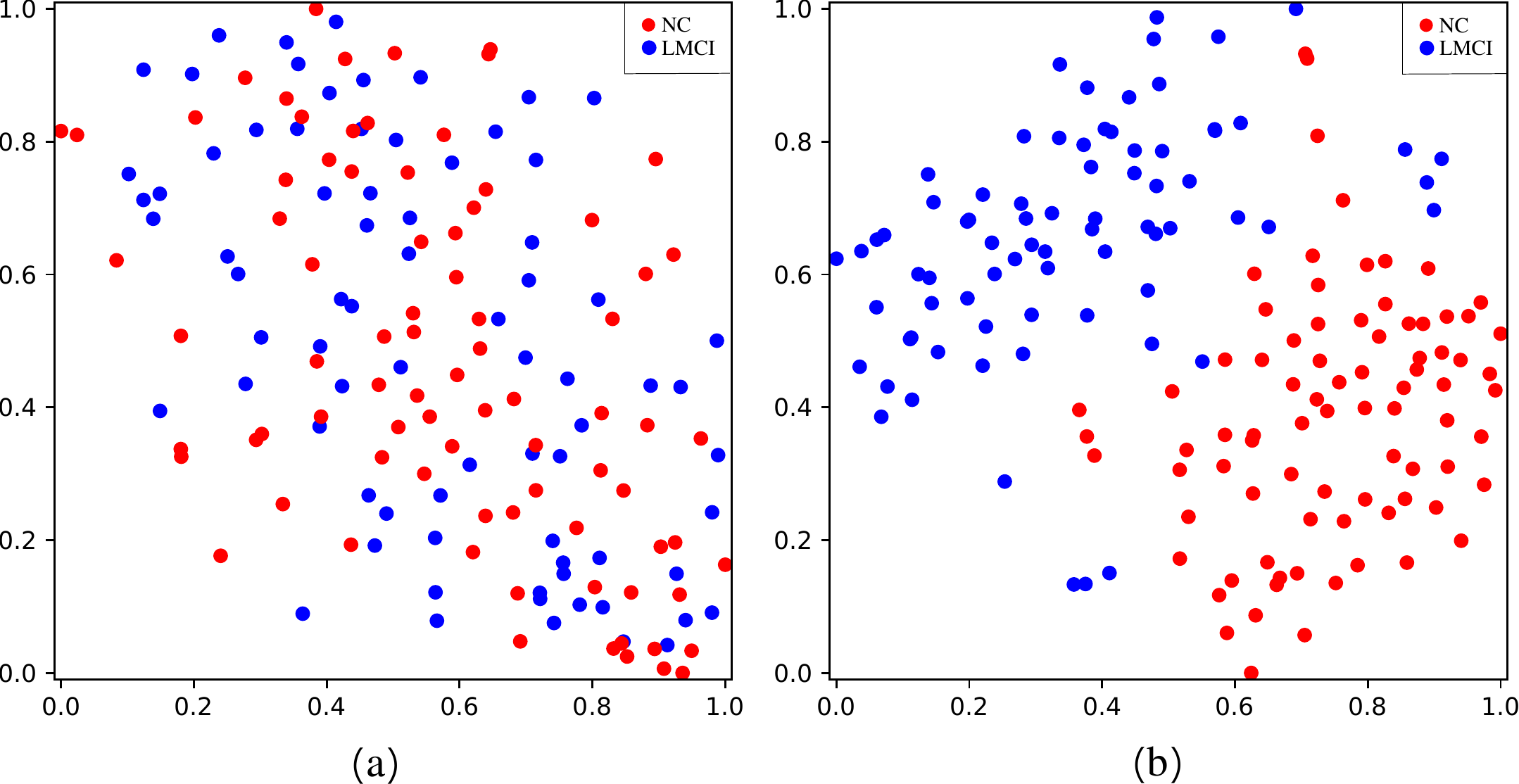}}
	\caption{The t-SNE visualization of the representations obtained by BSFL (a) without and (b) with the representation-fusing generator.}
	\label{fig15}
\end{figure}

In this study, the representation-fusing generator is important for unified brain network construction and analysis. To evaluate if the predictive UBN obtained by the generator is disease-related, the t-distributed stochastic neighbor embedding (t-SNE) tool ~\cite{ref_46_1} is used to display how the learned representations are arranged and if they are well separated. Fig.~\ref{fig15} shows the two-dimensional projection of the learned representations without and with the representation-fusing generator for NC vs. LMCI. The representations obtained by BSFL with the generator are well arranged by class information, while the ones obtained by BSFL without the generator are scattered and not badly separated. Thus, the BSFL can extract MCI-related features and capture complementary information between modalities.

\begin{figure}[htbp]
	\centerline{\includegraphics[width=\columnwidth]{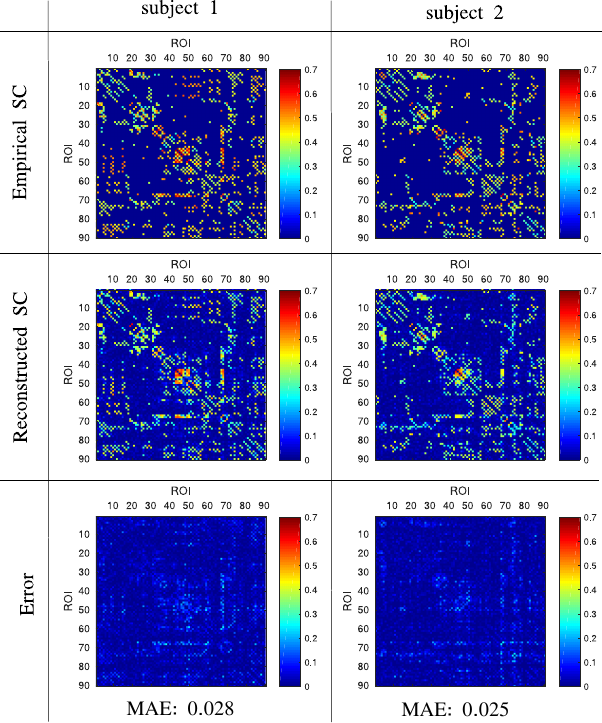}}
	\caption{Visualization of the empirical SC, the reconstructed SC, and the reconstructed error from two representative subjects.}
	\label{fig16}
\end{figure}

\begin{figure*}[htbp]
	\centering
	\includegraphics[scale=0.59]{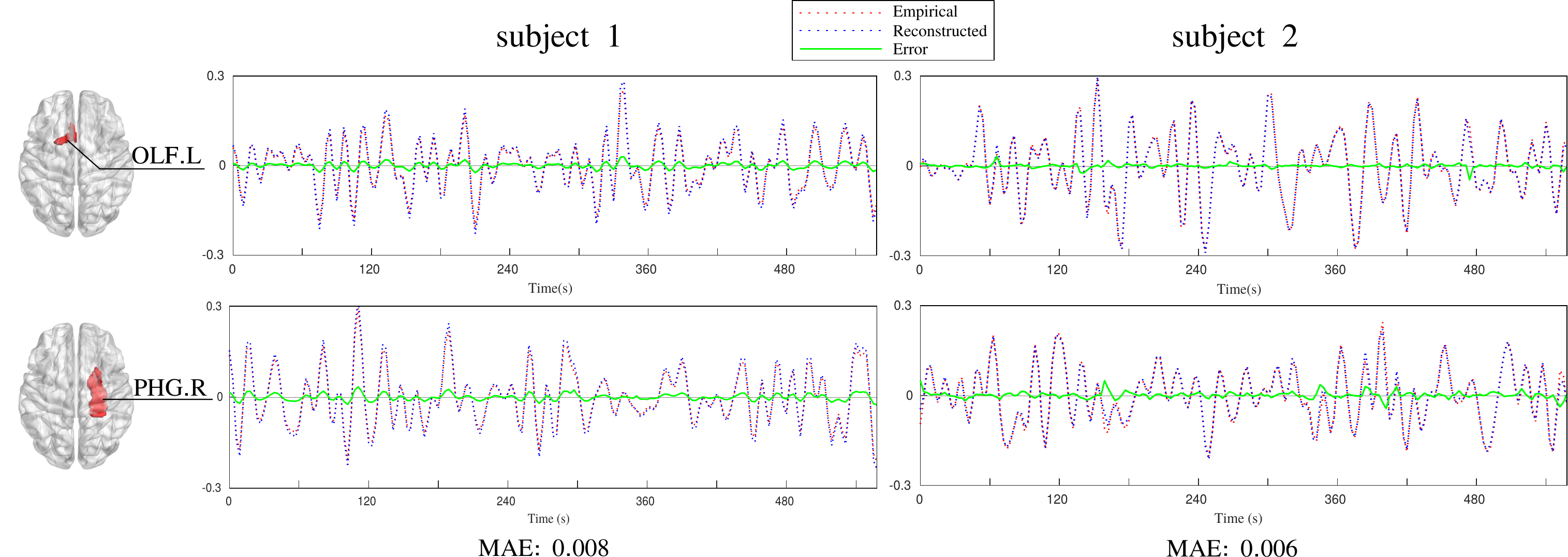}
	\caption{Visualization of the empirical functional time series, reconstructed time series, and the reconstructed error from two subjects on the two ROIs.}
	\label{fig17}
\end{figure*}

The structural and functional decoders are used to reconstruct the empirical SC and functional time series from the decomposed representations. And these representations are fused to generate unified brain networks for disease analysis. Therefore, the reconstruction process greatly influences the fusion effect and downstream tasks. To demonstrate the reconstruction performance, two representative subjects are selected from LMCI and NC respectively, and visualize the reconstructed SC and functional time series. As shown in Fig.~\ref{fig16}, the structural decoder can well rebuild the empirical SC. The mean absolute error (MAE) is used to evaluate the reconstruction quality. Fig.~\ref{fig17} shows the reconstructed time series on the two ROIs. From the above analysis, it can be seen that the two decoders perform well in the proposed model.

\subsection{Comparison with Clinical Studies}
Effective disease-related biomarkers are essential for clinicians to early diagnose neurodegenerative disease and develop treatments in delaying disease progression. The proposed model can output important ROIs and potential biomarkers in MCI analysis. In this section, comparing our results with clinical studies will be investigated. Considering the prediction tasks and the two-sample t-test, 10 top overlapping ROIs (ORBinf.R, OLF.L, OLF.R, SFGmed.R, PHG.R, AMYG.L, SOG.R, FFG.L, PCL.R, PUT.R) can be obtained in the identified brain regions in Table~\ref{Table2}. These ROIs are highly correlated with MCI and can be found in previous studies~\cite{ref_48,ref_49}. For example, the olfactory cortex plays an important role in translating everyday experiences into lasting episodic and working memory, which is preferentially attacked at the early stage of Alzheimer's disease. The parahippocampal gyrus also contributes to memory storage, and its structural damage can cause abnormal emotional and cognitive behavior. The paracentral lobule can recognize spatial relationships, in which patients with MCI showed atrophy in the parietal lobe. 

In the prediction of abnormal connections, the number of reduced connections increases from SMC to LMCI when compared with NC, while the number of increased connections also showed the same pattern. It indicates that these disease stages behave as a compensatory mechanism~\cite{ref_50} to make sure the brain function normally. Besides, the obtained important abnormal connections partly agree with clinical findings. For example, some increased brain connections have been identified by clinical discoveries~\cite{ref_51}, including HIP.L-ANG.L, HIP.L-ANG.R, PHG.R-TPOsup.R, PHG.R-TPOmid.R. And some reduced connections have been verified between left posterior cingulate gyrus and left hippocampus, and between right hippocampus and left middle occipital gyrus. Furthermore, the changes in brain connection strength increase or decrease dramatically as the disease progresses. The increased connections likely appear on the same brain hemisphere, while reduced connections are found across the brain hemispheres. It may be illustrated by the previous work that the increased connections have shorter distances compared with the reduced connections~\cite{ref_52}.

\begin{table}[t]
	\renewcommand\arraystretch{1.2}
	\caption{The top ten MCI-related ROIs verified by the clinical studies.}\label{Table2}
	\begin{center}
		\begin{tabular}{cccc}
			\toprule
			\textbf{ROI index} & \textbf{ROI name} & \textbf{Location} & \textbf{Verification}\\
			\midrule
			16 & ORBinf.R  & Frontal lobe & Hu $et\ al.$~\cite{ref_49_01}\\
			21 & OLF.L  & Frontal lobe    & Li $et\ al.$~\cite{ref_49_02}\\
			22 & OLF.R  & Frontal lobe    & Li $et\ al.$~\cite{ref_49_02}\\
			24 & SFGmed.R  & Frontal lobe & Yu $et\ al.$~\cite{ref_49_03}\\
			40 & PHG.R  & Temporal lobe   & Duan $et\ al.$~\cite{ref_49_04}\\
			41 & AMYG.L  & Temporal lobe  & Mihaescu $et\ al.$~\cite{ref_49_05}\\
			50 & SOG.R  & Occipital lobe  & Wang $et\ al.$~\cite{ref_49_06}\\
			55 & FFG.L  & Temporal lobe   & Hu $et\ al.$~\cite{ref_49_01}\\
			70 & PCL.R  & Parietal lobe   & Tang $et\ al.$~\cite{ref_49_07}\\
			74 & PUT.R  & Subcortical area & Wang $et\ al.$~\cite{ref_49_08}\\
			\bottomrule
		\end{tabular}
	\end{center}
\end{table}

\subsection{Limitation and Future Work}
Although the proposed model has achieved good performance in MCI analysis, there are two major limitations. One limitation is that the brain regions defined by the AAL template are a little coarse to represent the whole brain by constructing the unified brain network. The subtle changes between brain regions could not be detected. More fine ROI-based templates will be considered. Another limitation is that the subject size in this study is relatively small. It is better to control the number of parameters in the proposed model. Also, a much larger dataset should be tested to prove the effectiveness of the proposed model in the future.

\section{Conclusion}
In this paper, the proposed BSFL model is proposed to predict brain network abnormalities during MCI progression by combining fMRI and DTI. With the guidance of prior knowledge, the designed transformer can automatically extract the local and global connectivity features throughout the brain. The decomposed variational graph autoencoders decompose the feature space into unique and uniform spaces, and the uniform-unique contrastive loss is utilized to further improve the decomposition's effectiveness. The representation-fusing generator is utilized to fuse the decomposed representations to generate MCI-related connectivity features. The extensive experiments on the ADNI database demonstrate the proposed model's effectiveness compared with other competitive methods. Furthermore, some MCI-related brain regions and abnormal connections identified in our results also show the proposed model's reliability. Altogether, the proposed model is promising in reconstructing unified brain networks for brain disease analysis and providing potential connection-based biomarkers during the degenerative process of MCI.

\end{document}